\documentclass [epj,twocolumn,a4paper,showpacs]{revtex4}

\usepackage{graphics}

\begin {document}

\title {Spontaneous spin polarization in doped semiconductor quantum wells}

\author {L.\ O.\ Juri}
\email[e-mail: ]{luisjuri@jdcomp.com.ar}
\author {P.\ I.\ Tamborenea}
\email[e-mail: ]{pablot@df.uba.ar}

\affiliation{Department of Physics ``J.\ J.\ Giambiagi'',
University of Buenos Aires, Ciudad Universitaria, Pab.\ I,
C1428EHA Buenos Aires, Argentina}

\begin{abstract}
We calculate the critical density of the zero-temperature,
first-order ferromagnetic phase transition in $n$-doped
GaAs/AlGaAs quantum wells. We find that the existence of the
ferromagnetic transition is dependent upon the choice of well
width. We demonstrate rigorously that this dependence is governed
by the interplay between different components of the exchange
interaction and that there exists an upper limit for the well
width beyond which there is no transition. We predict that some
narrow quantum wells could exhibit this transition at electron
densities lower than the ones that have been considered
experimentally thus far. We use a screened Hartree-Fock
approximation with a polarization-dependent effective mass, which
is adjusted to match the critical density predicted by Monte Carlo
calculations for the two-dimensional electron gas.
\end{abstract}

 \maketitle

\section{\label{sec:intro}Introduction}

The interacting electron gas is one of the fundamental systems of
physics. However, in spite of a long tradition of study, the
subject still has many open basic questions. Notably, the issue of
the existence of a ferromagnetic transition at low density has not
been settled \cite{bloch}. Coulomb correlations play a central
role in the low-density regime, and taking them into account
theoretically (i.e., going beyond Hartree-Fock) is unfortunately
notoriously difficult. This problem has been most reliably tackled
with numerically intensive Monte Carlo (MC) techniques
\cite{ceper,tancep,att,ortiz}. For the two-dimensional electron
gas (2DEG), MC calculations indicate that, at $T=0$, a first-order
phase transition takes place at a certain critical value $r_{sc}$
of the dimensionless average separation between electrons $r_{s}
\equiv 1/\sqrt{\pi N_{s}}a^{*}_{B}$, where $N_{s}$ is the surface
density and $a^{*}_{B}$ is the effective Bohr radius in the
embedding medium ($a^{*}_{B}=98.7$\AA\ for GaAs).

The most widely used methods in MC calculations \cite{foulkes} are
the variational Monte Carlo (VMC), which predicts
\cite{ceper,tancep} a first-order phase transition at
$r_{sc}=13\pm 2$ ($N_{sc}=1.9\times10^{9}$cm$^{-2}$), and
fixed-node diffusion Monte Carlo (FN-DMC) with which $r_{sc}=25$
($N_{sc}=5.2\times10^{8}$cm$^{-2}$) has been found \cite{att}. The
VMC method uses a stochastic integration to evaluate the
ground-state energy for a given trial wave function. The other
method, which provides lower and more accurate ground-state
energies, uses a projection technique to enhance the ground-state
component of a trial wave function. In addition, in the FN-DMC
method implemented in reference \cite{att}, backflow correlations
\cite{kwon} are included in the Slater determinant of the trial
wave function, i.e. correlations are taken into account at the
starting point of the process, for each polarization.

Of course, the ideal, purely two-dimensional electron gas cannot
exist in nature. Instead, experimentally one studies
quasi-two-dimensional electron gases (quasi-2DEG) like the ones
formed in modulation-doped semiconductor quantum wells (QWs)
\cite{ando}. To the best of our knowledge, no Monte Carlo studies
comparable to the ones mentioned above have been done for
quasi-2DEG, but the ferromagnetic transition in QWs has been
studied theoretically in the frame of the local-spin-density
approximation \cite{tamb}. The critical densities predicted with
that technique exceed by far the density interval given by MC for
the 2DEG.

On the experimental front, the spin susceptibility has recently
been measured in GaAs/AlGaAs superlattices \cite{zhu}, with
electron densities as low as $1.7\times10^{9}$cm$^{-2}$. In spite
of the fact that this value falls into the density range predicted
for a transition by the 2DEG-MC calculations, no transition was
observed.

In this work, we study theoretically the possibility of a
first-order transition at $T=0$ for the quasi-2DEG confined in
GaAs/AlGaAs QWs as a function of the well width. We find that the
width and the depth of the well play a crucial role in the
existence of the transition. We prove rigorously the origin of the
pronounced dependence of the transition density on the well width
and predict the existence of an upper limit for this parameter
beyond which the polarized phase is energetically unfavorable. We
find that the transition should happen at electron densities lower
than those attained experimentally so far \cite{zhu}, for an
optimum value of the well width, which we provide below. In our
calculations we use a screened Hartree-Fock approximation scheme
that includes a polarization-dependent effective mass which is
introduced in order to take into account more accurately the
effects of Coulomb correlation inside the well. This novel
approach has the virtue of allowing us to use the results of the
existing numerical Monte Carlo studies in two dimensions and
extend them in a reliable way to quasi-2DEG systems.

The paper is organized as follows. In Section \ref{subsec:formal}
we introduce the basic scheme of the screened Hartree-Fock
approximation and obtain the equations for the ground-state
energies for the 2DEG and the quasi-2DEG. In Section
\ref{subsec:results} we discuss the results of this approximation.
In Section \ref{sec:effmass} we describe the
polarization-dependent effective-mass approximation and we use it
in combination with the available 2DEG Monte Carlo data to make
predictions for QWs. We end in Section \ref{sec:conc} with a
summary of our conclusions.

\section{\label{sec:formal} Screened-Hartree-Fock theory with
polarization-independent effective mass}

\subsection{\label{subsec:formal} Formalism}

In a quasi-2DEG, the HF equation may be written \mbox{as
\cite{pk}}
\begin {eqnarray}
\lefteqn{\left[E_{n}^{(\zeta)}(k)-\frac{\hbar^{2}k^{2}}{2m_{b}^{\ast}}\right]
                \Phi_{nk}^{(\zeta)}(z)=}  \nonumber   \\
 && \left[-\frac{\hbar^{2}}{2m_{b}^{\ast}}\frac{d^{2}}{dz^{2}}+V_{ext}(z)
                 +V_{sc}^{(\zeta)}(z)\right]\Phi_{nk}^{(\zeta)}(z) \nonumber \\
 && -\frac{2\pi e^{2}}{\varepsilon}\frac{1}{A}\int dz'\sum_{n' occup.}
                 \sum_{|\textbf{k}'|<k^{(\zeta)}_{Fn'}}
                 \frac{e^{-|\textbf{k}-\textbf{k}'||z-z'|}}{|\textbf{k}-
                 \textbf{k}'|} \nonumber \\
 && \times \Phi^{(\zeta)\ast }_{n'k'}(z')\Phi_{nk}^{(\zeta)}(z')\Phi_{n'k'}^{(\zeta)}(z),
\label{eq:hf}
\end {eqnarray}
where $\Phi_{nk}^{(\zeta)}(z)$ are the $n$th subband eigenstates
and $E_{n}^{(\zeta)}(k)$ the corresponding eigenenergies,
$m_{b}^{\ast}=0.067 m_{e}$ is the effective mass ($m_{e}$ being
the electron rest mass), $\varepsilon=12.5$ is the dielectric
constant, $e$ is the electron charge, $A$ is the crystal area, and
\textbf{k} is the in-plane wave vector. In all our calculations we
take the z-axis as the growth direction of the heterostructure.
The self-consistent potential $V_{sc}^{(\zeta)}(z)$ is obtained by
integration of the Poisson equation and is expressed as
\begin {equation}
V^{(\zeta)}_{sc}(z)=-\frac{4\pi
e^{2}}{\varepsilon}\left(\int_{0}^{z}dz'(z-z')n^{(\zeta)}(z')-\frac{N_{s}}{2}z\right),
\label{eq:vself}
\end {equation}
where the $\zeta$-dependent electron density is
\begin {equation}
n^{(\zeta)}(z)=\frac{2-\zeta}{2\pi}\sum_{n\
 occup.}\int_{0}^{k^{(\zeta)}_{Fn}}kdk|\Phi^{(\zeta)}_{n}(z,k)|^{2}.
\label{eq:dens}
\end {equation}
Here, $N_{s}$ is the doping sheet density and the external
potential $V_{ext}(z)$ is the sum of the confinement potential of
the heterostructure plus the electrostatic potential generated by
the ionized donors (located symmetrically). The Fermi level
$k_{Fn}^{(\zeta)}$ for each subband satisfies $2\pi
(1+\zeta)N_{s}=\sum_{n}k_{Fn}^{(\zeta)2}$. The spin polarization
index only takes the values $\zeta=0$ and $\zeta=1$ since, at
$T=0$, no stable partially-polarized phases are possible in 2DEG
\cite{att,dha1}. In contrast, recent calculations in 3DEG show
that the transition is not of first order, but rather a continuous
one involving partial spin-polarization states \cite{ortiz}. Here
we make the hypothesis that the quasi-2DEG behaves like the 2DEG
provided that the well width remains sufficiently small.

We solve equation (\ref{eq:hf}) following a method similar to that
developed in reference \cite{pk}. We expand the eigenfunctions
$\Phi_{nk}^{(\zeta)}(z)$ in the single-electron QW-basis functions
$\{\phi_{n}(z),\epsilon_{n}\}$, i.e.
\begin {equation}
\Phi_{nk}^{(\zeta)}(z)=\sum_{p}a_{pn}^{(\zeta)}(k)\phi_{p}(z).
\label{eq:expand}
\end {equation}
This enables us to write the eigenvalue equation
\begin {equation}
\hat{H}^{(\zeta)}(k)\vec{b}_{n}^{(\zeta)}(k)=E_{n}^{(\zeta)}(k)\vec{b}_{n}^{(\zeta)}(k),
\label{eq:eigeneq}
\end {equation}
with
$\vec{b}_{n}^{(\zeta)}(k)=(a_{1n}^{(\zeta)}(k),...,a_{pn}^{(\zeta)}(k),...)^{T}$.
The matrix elements of the Hamiltonian operator
$\hat{H}^{(\zeta)}(k)$ are
\begin {eqnarray}
H_{tp}^{(\zeta)}(k)=\left[\epsilon_{p}+\frac{\hbar^{2}k^{2}}{2m_{b}^{\ast}}\right]
\delta_{tp}+\langle\phi_{t}|V_{sc}^{(\zeta)}|\phi_{p}\rangle  \nonumber  \\
-V^{(1)(\zeta)}_{tp}(k)-V^{(2)(\zeta)}_{tp}(k), \label{eq:htp}
\end {eqnarray}
\begin {eqnarray}
V^{(1)(\zeta)}_{tp}(k)=\frac{e^{2}}{\varepsilon}\sum_{n'}
\int_{0}^{k^{(\zeta)}_{Fn'}}k'dk'\sum_{qr}G_{tr,qp}(k,k')   \nonumber   \\
\times a_{qn'}^{(\zeta)}(k')a_{rn'}^{(\zeta)}(k'),
\label{eq:v1tp}
\end {eqnarray}
\begin {eqnarray}
V_{tp}^{(2)(\zeta)}(k)=\frac{e^{2}}{\varepsilon}\sum_{n'}
\int_{0}^{k^{(\zeta)}_{Fn'}}k'dk' a_{pn'}^{(\zeta)}(k')a_{tn'}^{(\zeta)}(k')    \nonumber   \\
\times \frac{2}{\pi}\int_{0}^{\pi/2}\frac{d\varphi}
             {\sqrt{(k+k')^{2}-4kk'\sin^{2}\varphi}+q^{(\zeta)}_{s}},
\label{eq:vs23tp}
\end {eqnarray}

\begin {eqnarray}
G_{tr,qp}(k,k')=\int\int dzdz'\phi^{\ast }_{t}(z)\phi_{r}(z)
\phi^{\ast}_{q}(z')\phi_{p}(z')  \nonumber   \\
\times\int_{0}^{2\pi}\frac{d\varphi}{2\pi}
\frac{e^{-|\textbf{k}-\textbf{k'}||z-z'|}-1}{|\textbf{k}-\textbf{k'}|}.
\label{eq:gtrqp}
\end {eqnarray}
\begin{figure}
 \includegraphics[20mm,10mm][70mm,65mm]{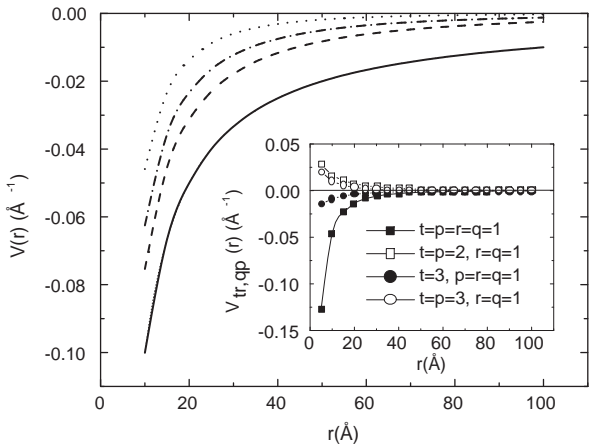}

 \caption{\label{fig3} Different Coulomb potentials.
The solid line corresponds to the unscreened Coulomb potential
$V(r)=1/r$. The dashed (dot-dashed) curve corresponds to the
Thomas-Fermi screened Coulomb potential for the polarized
(unpolarized) case. The dotted curve represents the potential
$V(r,d_{W})$, Fourier transform of equation (\ref{eq:gtrqp}) (with
respect to $q=|\textbf{k}-\textbf{k'}|$) for an infinite QW of
$d_{W}=100$\AA, for $t=r=q=p=1$. Inset: the Fourier transform of
equation (\ref{eq:gtrqp}) for an infinite QW of $d_{W}=100$\AA\
for different values of $t,r,q,p$.}
\end{figure}

We note that equations from reference \cite{pk} have a number of
misprints which are corrected here \cite{com1}.

To introduce screening in the HF approximation we note from
equations (\ref{eq:v1tp}) and (\ref{eq:vs23tp}) that
$V^{(2)(\zeta)}_{tp}(k)$ is the only term present in the pure 2D
case and describes the long-range in-plane Coulomb interaction.
Thus, we have dressed the interaction line of the exchange diagram
\cite{mano} by replacing the bare Coulomb potential
$V(q=|\textbf{k}-\textbf{k'}|)=(2\pi e^{2}/\varepsilon)(1/q)$ in
equation (\ref{eq:vs23tp}) with the statically screened Coulomb
potential
\begin {equation}
V^{(\zeta)}_{s}(q)=\frac{2\pi
e^{2}}{\varepsilon}\frac{1}{q+q^{(\zeta)}_{s}}, \label{eq:vscr}
\end {equation}
where $q^{(\zeta)}_{s}=(2-\zeta)/a^{\ast}_{B}$ is the
$\zeta$-dependent Thomas-Fermi wave number for the 2DEG.
Unfortunately, the Fourier transform in 2D real space of equation
(\ref{eq:vscr}) cannot be obtained analytically but its decay at
large $r$ is found to be \cite{davies}
\begin {equation}
V_{s}^{(\zeta)}(r)=\frac{e^{2}}{\varepsilon}\frac{1}{q^{2(\zeta)}_{s}r^{3}}.
\label{eq:vrs2d}
\end {equation}

On the other hand, $V^{(1)(\zeta)}_{tp}(k)$ (\mbox{Eq.\
(\ref{eq:v1tp})}) which arises from the intrinsic inhomogeneity of
charge distribution in a quasi-2DEG, represents an interaction of
short range. To see this, we consider the Fourier transform in 2D
real space
\begin {equation}
V(r,|z-z'|)=\frac{e^{2}}{\varepsilon}\left(\frac{1}{\sqrt{r^{2}+|z-z'|^{2}}}-\frac{1}{r}\right),
\label{eq:vrscr}
\end {equation}
of the potential $V(q,z-z')=(2\pi
e^{2}/\varepsilon)(e^{-q|z-z'|}-1)/q$ contained in equation
(\ref{eq:gtrqp}). It can be shown that $V(r,|z-z'|)$ is of short
range in the plane \cite{vasko} and that at large $r$ it can be
approximated as
\begin {equation}
V(r,d_{W})=-0.032\frac{e^{2}}{\varepsilon}\frac{d^{2}_{W}}{r^{3}},
\label{eq:vrscrapp}
\end {equation}
where we calculate the $z$ and $z'$ integration of
\mbox{$|z-z'|^{2}$} for an \textit{infinite} QW of well width
$d_{W}$ setting the subband indexes equal to one. From equations
(\ref{eq:vrs2d}) and (\ref{eq:vrscrapp}) we note that both
potentials decay at the same rate at large $r$ and that
$V(r,d_{W})\approx 0.128 V_{s}^{(0)}(r)$ and $V(r,d_{W})\approx
0.032 V_{s}^{(1)}(r)$ for $d_{W}= a^{*}_{B}$.

We show in Figure \ref{fig3} that the potential defined by
equation (\ref{eq:vrscr}), suitably integrated over $z$ and $z'$
(dotted line), has a range that is shorter than the range of the
Thomas-Fermi screened Coulomb potential for the polarized (dashed
line) and unpolarized (dot-dashed line) cases, for $t=r=q=p=1$. In
the inset we show that this case dominates over all others (we
only show a few relevant examples).

 In what follows we assume that only the first subband is occupied,
and therefore the summations over the subband index may be
omitted.

 In solving the eigenvalue equation by iteration, we consider that
self-consistency is achieved at the $l$th step when
$|a^{(\zeta)(l)}_{pn}(k)-a^{(\zeta)(l-1)}_{pn}(k)|/|a^{(\zeta)(l-1)}_{pn}(k)|<10^{-4}$
for all $p,n$ and $k$. We obtain the set of eigenfunctions and
eigenenergies $\{\Phi^{(\zeta)}_{nk}(z),E^{(\zeta)}_{n}(k)\}$ for
$\zeta=0$ and $\zeta=1$, which allows us to write the ground-state
energy per particle

\begin {equation}
E^{(\zeta)}_{HF}=\frac{2-\zeta}{4\pi
N_{s}}\int_{0}^{k^{(\zeta)}_{F}}kdk\left[E^{(\zeta)}_{1}(k)+
\tilde{\epsilon}^{(\zeta)}_{1}(k)+\frac{\hbar^{2}k^{2}}{2m_{b}^{\ast}}\right],
\label{eq:ehf}
\end {equation}
with
\begin {equation}
\tilde{\epsilon}^{(\zeta)}_{1}(k)=
\sum_{n}\epsilon_{n}|a^{(\zeta)}_{n1}(k)|^{2}. \label{eq:etilde}
\end {equation}

In order to check our quasi-2DEG calculations in QWs when $d_{W}$
tends to zero, we shall compare our results to the 2DEG screened
HF case \cite{fw}
\begin {equation}
E^{(\zeta)}_{HF2D}=\frac{e^{2}}{2a^{\ast}_{B}\varepsilon}
\left\{\frac{1+\zeta}{r_{s}^{2}}-\frac{4}{\pi
r_{s}}[2\zeta + \sqrt{2} (1-\zeta )]\textbf{I}(x_{\zeta})\right\},
\label{eq:ehf2d}
\end {equation}
where $x_{\zeta}=\frac{1}{4}[\zeta + 2\sqrt{2} (1-\zeta )]r_{s}$
is the polarization-dependent Thomas-Fermi wave number divided by
$2k^{(\zeta)}_{F}$ and
\begin {equation}
\textbf{I}(x_{\zeta})=\int_{0}^{1}\frac{xdx}
{x+x_{\zeta}}[\arccos(x)-x\sqrt{1-x^{2}}].
\label{eq:integ}
\end {equation}

\subsection{\label{subsec:results} Results}

In this section we present the results obtained with the screened
Hartree-Fock approximation with polarization-independent effective
masses introduced in the previous section.
\begin{figure}
 \includegraphics[20mm,10mm][70mm,65mm]{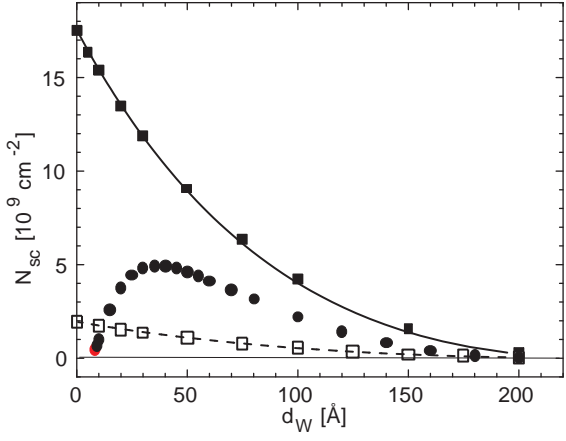}
 \caption{\label{fig1} Well-width dependence of the critical density in the
 screened HF approximation.
 Solid squares correspond to \textit{infinite} quantum wells
 and solid circles to \textit{finite} QWs with well height $V_{b}$=247 meV.
 Open squares correspond to \textit{infinite} QWs with
 polarization-dependent effective masses with a constant ratio
 $f \equiv m^{\ast}_{1}/m^{\ast}_{0}=0.65$. The solid and dashed lines are obtained with
 equations (\ref{eq:trans}) and (\ref{eq:ftrans}) (with $f=0.65$), respectively,
 for \textit{infinite} QWs.}
\end{figure}

In Figure \ref{fig1}, we plot the critical density for infinite
QWs (in which there is no barrier penetration) in the screened HF
approximation as a function of well width $d_{W}$ (solid squares).
The limiting point at $d_{W}=0$ (2DEG) was calculated with
equation (\ref{eq:ehf2d}) and the remaining points with equation
(\ref{eq:ehf}). The excellent match between these two different
equations reflects the correctness of our derivations and
calculations \cite{com1}. We see that the critical density for the
\textit{infinite} QWs is a rapidly decreasing function of $d_{W}$.
This result, which looks superficially correct on the basis of a
gradual 2D to 3D transition, can be explained rigorously in terms
of the interplay between different components of the exchange
interaction in a quasi-2DEG, as follows. We found that the
coefficients $a^{(\zeta)}_{p1}(k)$ (\mbox{Eq.\ (\ref{eq:expand})})
change very little with $k$ and $a^{(\zeta)}_{11}(k)\approx 1$
whereas $a^{(\zeta)}_{p1}(k)\ll 1$ for $p>1$. This result is a
direct consequence of the small-density condition since, if this
condition is met, the ground-state wave function must retain the
shape of its one-electron counterpart in a QW. Thus,
$V_{11}^{(2)(\zeta)}(k)$ (\mbox{Eq.\ (\ref{eq:vs23tp})}) is
positive and it is the leading matrix element, yielding a negative
(see \mbox{Eq.\ (\ref{eq:htp})}) and $d_{W}$-independent
contribution to the exchange energy (pure 2D case). In contrast,
the matrix elements $V^{(1)(\zeta)}_{tp}(k)$ are slowly varying
functions of $k$ since we are studying small-width and low-density
heterostructures. If these conditions are fulfilled, we can expand
the exponential in equation (\ref{eq:gtrqp}) up to first order
since the exponent satisfies the following inequality:
\begin {equation}
|\textbf{k}-\textbf{k'}||z-z'|\leq
2k^{(\zeta)}_{F}d_{W}=2\sqrt{2(1+\zeta)}\frac{1}{r_{s}}\frac{d_{W}}{a^{*}_{B}}
\leq 0.3, \label{eq:aprox1}
\end {equation}
provided that $r_{s}\geq 13$ (lower MC limit for the transition)
and the well width $d_{W}\leq a^{*}_{B}$ ($=98.7$\AA). Then we
perform the integration over $z$ and $z'$ (\textit{infinite} QWs)
yielding $G_{11,11}=-0.207 d_{W}$. Thus, $V_{11}^{(1)(\zeta)}$ is
negative, proportional to $d_{W}$ and $k$-independent. The matrix
elements $V_{tp}^{(1)(\zeta)}$ for indices with opposite parity
are always zero since due to the symmetry of the wave functions we
have $G_{t1,1p}=0$. On the other hand, we have evaluated
$G_{t1,1p}$ for indices with equal parity obtaining that these are
considerably smaller than $G_{11,11}$ rendering
$V_{tp}^{(1)(\zeta)}$ practically diagonal. Also,
$\langle\phi_{t}|V_{sc}^{(\zeta)}|\phi_{p}\rangle \approx 0$ (see
\mbox{Eq.\ (\ref{eq:htp})}) for the non-diagonal elements since,
due to the low-density condition, $V_{sc}^{(\zeta)}$ must be a
very slowly varying function of $z$. Thus, $H_{tp}^{(\zeta)}(k)$
is almost diagonal, being its first eigenenergy
\begin {equation}
E^{(\zeta)}_{1}(k)\approx\epsilon_{1}+\frac{\hbar^{2}k^{2}}{2m_{b}^{\ast}}+\langle\phi_{1}|V^{(\zeta)}_{sc}|\phi_{1}\rangle
-V^{(1)(\zeta)}_{11}-V^{(2)(\zeta)}_{11}(k).   \label{eq:e1}
\end {equation}

With this expression for $E^{(\zeta)}_{1}(k)$ we can obtain an
approximate equation for $N_{sc}$. We only need to make two easily
justified additional approximations. From the behavior of the
coefficients $a^{(\zeta)}_{p1}(k)$, i.e. $a^{(1)}_{11}(k)\approx
a^{(0)}_{11}(k)\approx 1$ and $a^{(1)}_{p1}(k)\approx
a^{(0)}_{p1}(k)\approx 0$ for $p>1$ it can be seen (from
\mbox{Eq.\ (\ref{eq:etilde})}) that
$\tilde{\epsilon}^{(\zeta)}_{1}(k)\approx \epsilon_{1}$ and (from
\mbox{Eq.\ (\ref{eq:expand})}) $\Phi^{(\zeta)}_{1k}(z)\approx
\phi_{1}(z)$. Using the latter in equation (\ref{eq:dens}) we get
$n^{(\zeta)}(z)\approx
(2-\zeta)(1+\zeta)N_{s}\phi^{2}_{1}(z)/2=N_{s}\phi^{2}_{1}(z)$,
and therefore a $\zeta$-independent
$\langle\phi_{1}|V_{sc}|\phi_{1}\rangle$ (see \mbox{Eq.\
(\ref{eq:vself})}). By inserting equation (\ref{eq:e1}) in
equation (\ref{eq:ehf}) we obtain, after some algebra, the
following equation for the energy shift between both phases
\begin {equation}
E^{(1)}_{HF}-E^{(0)}_{HF}\approx \frac{\pi
N_{s}e^{2}a^{*}_{B}}{2\varepsilon}
\left[1+0.207\frac{d_{W}}{a^{*}_{B}}-4F(N_{s})\right],
\label{eq:eshift}
\end {equation}
where
\begin {eqnarray}
F(N_{sc})=\int_{0}^{1}xdx\int_{0}^{1}x'dx'   \nonumber    \\
\times\frac{2}{\pi}\int_{0}^{\pi/2}d\varphi
[g_{1}(N_{sc},x,x',\varphi) \nonumber  \\
-g_{0}(N_{sc},x,x',\varphi)], \label{eq:efe}
\end {eqnarray}
with
\begin {eqnarray}
g_{0}(N_{sc},x,x',\varphi)= \left[\sqrt{2\pi N_{sc}}a^{*}_{B}\right.  \nonumber   \\
\left.\times \sqrt{(x+x')^{2}-4xx'sin^{2}\varphi}+2\right]^{-1},
\label{eq:g0}
\end {eqnarray}
and
\begin {eqnarray}
g_{1}(N_{sc},x,x',\varphi)=\left[\sqrt{\pi N_{sc}}a^{*}_{B}\right.  \nonumber   \\
\left.\times\sqrt{(x+x')^{2}-4xx'sin^{2}\varphi}+q^{(1)}_{s}/q^{(0)}_{s}\right]^{-1},
\label{eq:g1}
\end {eqnarray}
where $x \equiv k/k^{(0)}_{F}$ in $g_{0}$ and $x \equiv
k/k^{(1)}_{F}$ in $g_{1}$. The same holds for $x'$ and $k'$.

Taking into account that the energy shift
$E^{(1)}_{HF}-E^{(0)}_{HF}$ must be zero at the transition
density, we may write the following equation that relates $d_{W}$
and $N_{sc}$
\begin {equation}
1+0.207\frac{d_{W}}{a^{*}_{B}}=4F(N_{sc}).    \label{eq:trans}
\end {equation}

Finally, to demonstrate that the transition density is a
decreasing function of the well width, we need to prove that
$F(N_{sc})$ is also a decreasing function. To see this, we note
firstly that $g_{1}>g_{0}$ for all values of its arguments making
the function $F(N_{sc})$ always positive allowing equation
(\ref{eq:trans}) to be solvable. Secondly, both $g_{0}$ and
$g_{1}$ are decreasing functions of $N_{sc}$ for all values of
$x,x',\varphi$ and it is straightforward to prove that $g_{1}$
decreases more rapidly than $g_{0}$ making $F(N_{sc})$ a
decreasing function. Thus, an increase of $d_{W}$ must be
accompanied with a decrease of $N_{sc}$ proving that the
monotonically decreasing dependence of $N_{sc}$ on $d_{W}$ is
governed by the competing action of the different components of
the exchange interaction: the in-plane component represented by
$F(N_{sc})$ and the out-of-plane term driven by $d_{W}$. The
behavior of $g_{0}$, $g_{1}$ and $F$ that we described can also be
recognized in the unscreened HF case indicating that the
dependence of $N_{sc}$ on $d_{W}$ is purely due to exchange. The
validity of our approximation can be verified in Figure \ref{fig1}
where the solutions of equation (\ref{eq:trans}) are depicted with
the solid line. This curve approaches very well the exact values
(solid squares) indicating that the approximations we made to
derive it are well justified. For low densities (high $r_{s}$) and
low well widths, the solid curve fits excellently the solid
squares since in these regimes the approximations we made become
exact. In the intermediate region ($d_{W}\approx 100$\AA) the
solid curve fits very well the solid squares.

A look at equation (\ref{eq:trans}) also indicates that there
exists an upper limit for the well width. In fact, due to the
decrease of $F(N_{sc})$ and since $F(0)$ is finite, it can be seen
from equations (\ref{eq:g0}) and (\ref{eq:g1}) that the following
relation holds
\begin {equation}
0.207\frac{d_{WL}}{a^{*}_{B}}=\frac{q^{(0)}_{s}}{q^{(1)}_{s}}-\frac{3}{2}.
\label{eq:dwl}
\end {equation}
Thus, it must be $d_{W}<d_{WL}=2.42a^{*}_{B}\approx 239$\AA\ to
allow the confined electrons to reach the polarized phase. This
new result is entirely due to the Thomas-Fermi screening and is
not present in the unscreened HF approximation since $F(0)$
diverges in this case. Furthermore, the ratio of the Thomas-Fermi
wave numbers for both polarizations must be
$q^{(0)}_{s}/q^{(1)}_{s}>3/2$ (in our case it is
$q^{(0)}_{s}/q^{(1)}_{s}=2$). Otherwise, no polarized state could
be possible in a QW. These key results indicate that the well
width plays a crucial role in the search for spontaneous spin
polarization in QWs \cite{tutuc}.

The implications of the existence, according to our calculations,
of an upper limit for the well width must be considered with some
care. If one attempts to reach a 3DEG system by increasing the
well width one would apparently fall into the paradox that no
transition is possible in 3DEG. This conclusion is incorrect for
two reasons. First, we must take into account that the 3DEG-MC
results are obtained in the jellium model, in which the positive
background is taken to be a uniform neutralizing static charge
distribution, whereas in our quantum-well calculations the
positive charges of the ionized donors are located far away from
the electron gas, which results in an important change in the
direct Coulomb energy. In other words, wide-enough quantum wells
and 3DEG jellium model must be considered as different systems.
Secondly, our calculation assumes that only one subband is
occupied (a valid assumption in narrow quantum wells at low
density) whereas any extrapolation of our conclusions to 3DEG
systems would have to contemplate necessarily occupation of many
subbands. Thus, we reach the conclusion that the most likely
scenario is that there is a phase transition in narrow quantum
wells, which disappears for intermediate well widths, and reenters
at wider well widths as expected when the 3DEG limit is
approached.

We now use the previous result to analyze the critical density for
\textit{finite} QWs, plotted in Figure \ref{fig1} with solid
circles. We set the height of the QWs to \mbox{$V_{b}=247$ meV}, a
typical experimental value \cite{bastard}. This curve exhibits a
non-monotonic dependence on the well-width showing a maximum for
$d_{W}\approx 35$\AA. Also we observe a general reduction of the
critical density with respect to the case of \textit{infinite}
QWs. This can be simply understood in terms of the previous result
(monotonically decreasing critical density for the
\textit{infinite} wells) and the penetration of the electron wave
function into the AlGaAs barriers; the latter causes the wave
function to spread beyond the nominal well width, effectively
``enlarging'' the well. As a consequence, for example, a
\textit{finite} QW of $d_{W}\approx 60$\AA\ has the same critical
density as that of an \textit{infinite} QW of $d_{W}\approx
100$\AA. In fact, the penetration depth $d_{B}=\hbar/
\sqrt{2m_{b}^{\ast}(V_{b}-E_{1})}$ increases when $E_{1}$ is
raised as $d_{W}$ is lowered \cite{vasko}. This effect produces an
inflection point at $d_{W}\approx 75$\AA\ and the mentioned
maximum at $d_{W}\approx 35$\AA\ due to the competition between
$d_{W}$ and $d_{B}$.

 Let us go back to the curve for \textit{infinite} QWs in Figure
\ref{fig1}. The limiting ($d_{W}=0$) value $N_{sc}=17.5\times
10^{9}$cm$^{-2}$ corresponds to $r_{sc}=4.32$, showing a sizable
increase with respect to the (unscreened) HF value $r_{sc}=2.01$
\cite{raja}. This increase, however, is not sufficient if we
consider the value $r_{sc}=13$ obtained in reference \cite{ceper}
using VMC. This indicates that a significant degree of Coulomb
correlation is being left out in the screened HF approximation.

\section{\label{sec:effmass}Polarization-dependent effective masses}

\subsection{\label{subsec:twodim} Two-dimensional case}

In order to go further and improve our treatment of Coulomb
correlation, we need an approximation scheme applicable to the
quasi-2DEG such that as $d_{W}$ tends to zero (pure 2DEG) the
critical density approaches the values predicted by MC
calculations \cite{ceper,tancep,att}. To achieve this, we
incorporate phenomenological polarization-dependent effective
masses $m_{0}^{\ast}$ (unpolarized) and $m_{1}^{\ast}$ (polarized)
in our formalism. Due to the lack of experimental data on
effective masses in GaAs/AlGaAs heterostructures for both
polarizations and that no calculations on polarized effective
masses exist in 2DEG, we resort to calculations of unpolarized
effective masses and ground-state energies in pure 2DEG
\cite{janmin,krako,ceper} to justify this procedure. Let us
summarize the conclusions of those
studies relevant in our context:\\
(a) Coulomb correlation increases the effective mass \cite{krako}.\\
(b) The absolute value of the correlation energy of the
unpolarized 2DEG
ground state is greater than its polarized counterpart \cite{ceper}.\\
(c) The absolute value of the correlation energy is greater in 2D
than in 3D (both unpolarized), leading to 2D effective masses
substantially larger than those of the 3D case at
equal $r_{s}$ \cite{janmin}.\\
(d) The correlation-energy shift between both phases in 2D is
greater than  the unpolarized correlation energy shift between 2D and 3D \cite{ceper}.\\
Making use of (a) and (b), with the supporting evidence of (c) and
(d), we conclude that \textit{ $m_{0}^{\ast}$ must be
substantially larger than $m_{1}^{\ast}$ at equal $r_{s}$.}
\begin{figure}
  \includegraphics[20mm,10mm][70mm,65mm]{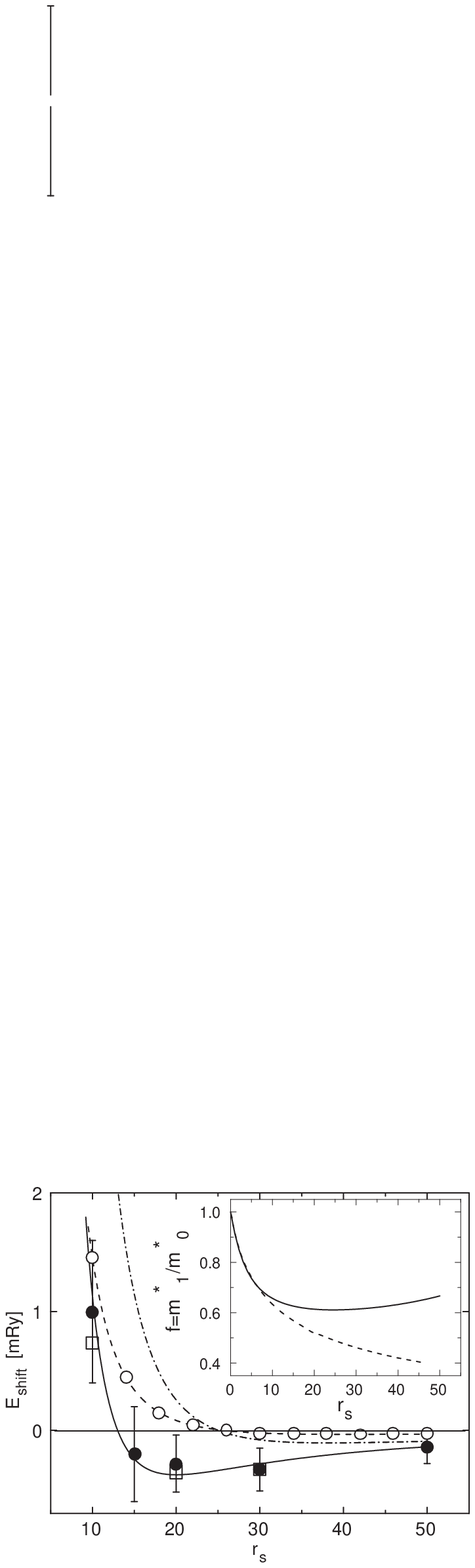}
  \caption{\label{fig2} Ground-state energy shift
   $E_{shift}=E_{HF2D}^{(1)}-E_{HF2D}^{(0)}$
   versus $r_{s}$. Solid circles represent the reported values in Table I
   of reference \cite{ceper}. The error bars denote the VMC standard errors.
   Open squares correspond to the values tabulated in Tables I and II for
   the VMC method in reference \cite{tancep}.
   No error bars are plotted for clarity. Open circles belong from
   reference \cite{att}.
   The solid and dot-dashed lines represent our calculations obtained with
   equation (\ref{eq:ehf2dm}) for polarization-dependent effective masses
   using $f=0.65$ and $f=0.49$ respectively. The dashed curve
   corresponds to the same calculations but using the values of $f$ that
   come from $f(r_{s})$ showed in the inset (dashed line). In the inset, solid and dashed
   lines correspond to the values of $f$ that fit the curves $E_{shift}$
   from reference \cite{ceper} and reference \cite{att} respectively.}
\end{figure}

By defining the ratio $f = m^{\ast}_{1} / m^{\ast}_{0}$ and
rewriting $x_{\zeta}=\frac{1}{4}[f\zeta + 2\sqrt{2} (1-\zeta
)]r_{s}$ we may write equation (\ref{eq:ehf2d}) as
\begin {eqnarray}
E^{(\zeta)}_{HF2D}=\frac{e^{2}m_{0}^{\ast}}{2a^{\ast}_{B}\varepsilon}\left\{ \frac{1+\zeta}{[(f-1)\zeta+1]r_{s}^{2}}\right.\nonumber  \\
\left. -\frac{4}{\pi r_{s}}[2\zeta +
 \sqrt{2}(1-\zeta)]\textbf{I}(x_{\zeta})\right\}.
\label{eq:ehf2dm}
\end {eqnarray}

 We observe from reference \cite{janmin} that $m^{\ast}_{0}\approx 1.2$
for $r_{s} > 5$ in the modified Hubbard approximation. That
approximation is an attempt at including correlation effects by
means of the introduction of the Thomas-Fermi wave number in the
so-called local-field correction factor. Since we have
incorporated screening correlations and HF effects within a
similar scheme, we take $m^{\ast}_{0}=1$ in equation
(\ref{eq:ehf2dm}) to avoid an overestimation of the effective mass
in the unpolarized phase.

 Using equation (\ref{eq:ehf2dm}), the lowest 2DEG-VMC
value, i.e. $r_{sc}=13$, is obtained with $f=0.65$. With this
value of $f$ we calculate
$E_{shift}=E^{(1)}_{HF2D}-E^{(0)}_{HF2D}$ and plot it versus
$r_{s}$ in Figure \ref{fig2} (solid line). Here we have taken
$e^{2}/2a^{\ast}_{B}\varepsilon=$ 1~Ry to compare our $E_{shift}$
against MC results. The solid circles correspond to the results
obtained in reference \cite{ceper} and the open squares are from
reference \cite{tancep}. We note the excellent agreement between
our curve and the MC points: by adjusting only one point our curve
meets all the points obtained in reference \cite{ceper,tancep}.
This agreement implies that the ratio between both effective
masses depends weakly on the density and supports the validity of
our assumption of a density-independent $f$ factor. This
conclusion is consistent with the fact that in the modified
Hubbard approximation the unpolarized effective mass is a slowly
varying function for $r_{s}>5$ \cite{janmin}. If this were also
the behavior of the polarized effective mass, we could conclude
that $f$ would be a slowly varying function of $r_{s}$.

 We now repeat the previous analysis but using the data of
Attaccalite \textit{et al.} \cite{att}. In that paper the authors
obtain $r_{sc}=25$ which, as we mentioned in the introduction, is
the highest value found in the literature for spontaneous spin
polarization in 2DEG at zero temperature. We find that equation
(\ref{eq:ehf2dm}) reproduces the value $r_{sc}=25$ when $f=0.49$.
In Figure \ref{fig2} (dash-dotted line) we plot the energy shift,
$E_{shift}=E^{(1)}_{HF2D}-E^{(0)}_{HF2D}$, versus $r_{s}$, for
this value of $f$. This curve does not fit well the data from
reference \cite{att} shown as open circles. Instead, we find that
the ratio $f$ now taken as a function of $r_{s}$ (dashed line in
the inset) fits very well the MC points calculated in reference
\cite{att} (open circles) when it is used in equation
(\ref{eq:ehf2dm}) (dashed line in Fig.\ \ref{fig2}. For
completeness, we show in the inset (solid line) the values of $f$
as a function of $r_{s}$ that fit the parametrization of
$E_{shift}$ obtained by Ceperley \cite{ceper}. This curve exhibits
a weak dependence on $r_{s}$ for $r_{s}>10$ giving support to our
initial assumption of a constant value $f=0.65$.

\subsection{\label{subsec:quasitwodim} Quasi-two-dimensional case}

\begin{figure}
 \includegraphics[20mm,10mm][70mm,65mm]{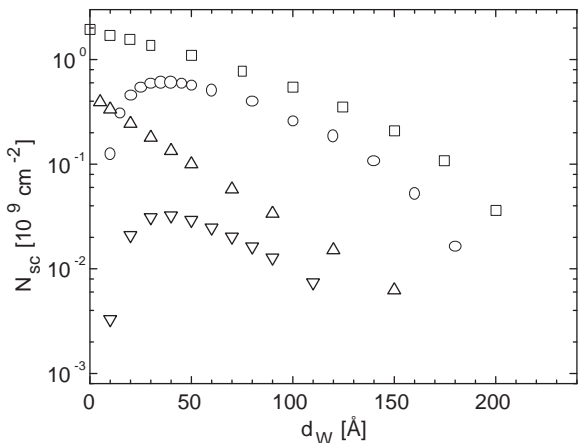}
 \caption{\label{fig4} Well-width dependence of the critical density in the
 screened HF approximation with polarization-dependent effective masses.
 Open squares (circles) correspond to \textit{infinite}
 (\textit{finite}) QWs  with a constant ratio $f \equiv m^{\ast}_{1}/m^{\ast}_{0}=0.65$.
 Up (down) triangles correspond to \textit{infinite} (\textit{finite}) QWs for a
 ratio $f$ which depends on $r_{s}$ (dashed line in the inset of Fig.\ \ref{fig2}).}
\end{figure}

We now apply the polarization-dependent effective-mass
approximation to the quasi-2DEG by means of a slight modification
in equation (\ref{eq:hf}). We note that the two effective masses
$m_{b}^{\ast}$ on both sides of equation (\ref{eq:hf}) belong to
different situations \cite{vasko}: the $m_{b}^{\ast}$ on the
l.h.s. represents the in-plane effective mass and therefore is
being affected by Coulomb correlations. In contrast, the
$m_{b}^{\ast}$ on the r.h.s. reflects the out-of-plane effective
mass of one electron moving in the z-direction governed mainly by
$V_{ext}(z)$ and $V_{sc}^{(\zeta)}(z)$, and thus not being
affected by Coulomb correlations, according to our discussion
about screening in Section \ref{subsec:formal}. Then we solve the
eigenvalue equation, equation (\ref{eq:hf}), and use equation
(\ref{eq:ehf}) for both polarizations incorporating the effective
masses $m^{\ast}_{0}=m^{\ast}_{b}$ and $m^{\ast}_{1}=f
m^{\ast}_{b}$ (the last one only in the in-plane terms). We plot
in Figure \ref{fig1}, with open squares, the results obtained for
\textit{infinite} QWs for $f=0.65$. We calculate the limiting
point at $d_{W}=0$ with equation (\ref{eq:ehf2dm}) and the
remaining points with equation (\ref{eq:ehf}) with the
above-mentioned replacement giving an excellent match. We show in
the Figure \ref{fig4}, in a logarithmic scale for the vertical
axis, the results for \textit{finite} QWs (open circles) and
\textit{infinite} QWs (open squares) where we have taken $f=0.65$.
Both curves exhibit the same general characteristics as in Figure
\ref{fig1} (solid squares and solid circles).

 Now we obtain an equivalent of equation (\ref{eq:trans})
by incorporating the ratio $f$ that multiplies the band mass
$m^{*}_{b}$ for the polarized phase, yielding
\begin {equation}
\frac{2}{f}-1+0.207\frac{d_{W}}{a^{*}_{B}}=4F(N_{sc},f),
\label{eq:ftrans}
\end {equation}
where $F(N_{sc},f)$ is the same as in equation (\ref{eq:efe}), but
$g_{1}$ now reads
\begin {eqnarray}
g_{1}(N_{sc},x,x',\varphi)=\left[\sqrt{\pi N_{sc}}a^{*}_{B}\right.  \nonumber   \\
\left. \times
\sqrt{(x+x')^{2}-4xx'sin^{2}\varphi}+fq^{(1)}_{s}/q^{(0)}_{s}\right]^{-1}.
\label{eq:g1f}
\end {eqnarray}

The polarization-dependent effective-mass approximation does not
change the previous result regarding the monotonically decreasing
dependence of $N_{sc}$ on $d_{W}$ since $f<1$. We show that the
solutions of the approximate equation (\ref{eq:ftrans}) (dashed
curve) depicted in Figure \ref{fig1} for $f=0.65$, fit perfectly
the exact values (open squares). We observe that since
$F(0,f)=\frac{1}{4}(\frac{2}{f}-\frac{1}{2})$, $d_{WL}$ does not
depend on $f$ (see Eq.\ (\ref{eq:ftrans})). Thus, $d_{WL}$ depends
on correlations, in our Thomas-Fermi model, via the ratio
$q^{(0)}_{s}/q^{(1)}_{s}$ and, consequently, the relation
$d_{WL}=2.42a^{*}_{B}$ is a universal one, i.e. it holds for QWs
of any material. We note that this interesting result does not
depend on the approximations we made to derive equation
(\ref{eq:ftrans}) since those become exact as $N_{s}$ tends to
zero.

 On the other hand, the location of the observed maximum remains
unchanged with respect to the $f=1$ case (solid circles in Fig.\
\ref{fig1}) in accordance to our initial assumption that screening
and correlations manifest only in the plane. Up (down) triangles
correspond to \textit{infinite} (\textit{finite}) QWs where we
have taken the ratio $f$ as the dashed curve in the inset of
Figure \ref{fig2} (Attaccalite \textit{et al.} \cite{att}). We
observe a drastic diminution of the transition densities in this
case for both \textit{infinite} and \textit{finite} QWs. We note
that the MC density interval for spontaneous spin polarization
mentioned in Section \ref{sec:intro}, appears notoriously shrunk
for the \textit{finite} QWs studied here. In fact, from Figure
\ref{fig4} we obtain a new density interval for the transition
densities in \textit{finite} QWs between
$N_{sc}=3.2\times10^{7}$cm$^{-2}$ and
$N_{sc}=6.1\times10^{8}$cm$^{-2}$. We take these values from the
transition densities at the maximum of the curves related to
FN-DMC (down triangles) and VMC (open circles) respectively.

 In reference \cite{zhu}, the spin susceptibility ($=m^{\ast}g^{\ast}$)
has been measured in a high quality 200-fold GaAs/AlGaAs
superlattice of 100\AA\ of GaAs wells and 30\AA\ barriers of
Al$_{0.32}$Ga$_{0.68}$As, with unprecedented low densities such as
$N_{s} = 1.7 \times 10^9 \mbox{cm}^{-2}$ ($r_{s}=13.9$) and no
transition was observed. According to what we have mentioned
above, this is not surprising. There are several possible reasons
for this negative result. We first note that the density used,
although low enough for a transition in the pure 2DEG, is clearly
too high considering the finite well width for \textit{finite}
QWs: for $d_{W}=100$\AA\ (\mbox{Fig.\ \ref{fig4}}), the electron
density achieved in reference \cite{zhu} is 6.5 times higher than
our critical density (open circles) which uses the ratio $f$ that
matches the 2DEG value from reference \cite{ceper} and 50 times
higher than the critical density (down triangles) which uses the
$r_{s}$-dependent ratio $f$ that matches the 2DEG energy shifts
$E_{shift}$ from reference \cite{att}. Also, due to the tunneling
of the electrons into the AlGaAs barriers, the superlattice acts
like a single, extremely wide QW. Furthermore, it is possible that
if the QW were sufficiently wide, the quasi-2DEG could lose its
two-dimensional characteristics, allowing for stable
partially-polarized phases like those possible in the 3DEG,
turning more difficult the detection of the transition. We note
that the effects of in-plane correlations combined with the finite
well widths and heights of QWs produce a drastic diminution of the
transition densities by a factor that ranges from 3 to 15
depending on which method VMC or FN-DMC turns out to be the best
tool to estimate the transition density in pure 2DEG. For the best
case, it should become necessary to achieve electron densities
lower that the ones studied experimentally thus far by a factor of
3 and by a factor of 53 in the worst case. Very different could be
the quasi-two-dimensional hole gas (quasi-2DHG) scenario since in
that system, high $r_{s}$ values such as $r_{s}\approx 80$ are
already attainable \cite{noh}. However, our theoretical
predictions about the critical transition density in n-doped
GaAs/AlGaAs QWs cannot be straightforwardly translated to
quasi-2DHG. The adaptation of our formalism to the problem with
holes is currently in progress.

Based on the insight gained from our calculations, we propose that
the optimal conditions for observing a ferromagnetic transition in multiple QWs are: \\
(a) well widths between $30$\AA\ and $50$\AA\,\\
(b) wide AlGaAs barriers between wells to prevent tunneling, and\\
(c) well height $V_{b}$ as large as possible to minimize
barrier-penetration effects.

In a recent experimental work, Gosh \textit{et al.} \cite{ghosh}
report a possible spontaneous spin polarization in mesoscopic
two-dimensional systems that is at odds with our findings. They
have used 2DEGs in Si $\delta$-doped GaAs/AlGaAs heterostructures
with densities as low as $N_{s}=5\times10^{9}$cm$^{-2}$
($r_{s}=7.6$) and the temperature was set at $T=40$ mK or
equivalently $T/T_{F}\approx 0.02$ since $T_{F}=2.3$ K at
$r_{s}=7.6$. Somewhat surprisingly, according to their
interpretation of the data, these authors found partial spin
polarization with $\zeta=0.2$. The authors attribute this partial
spin polarization to the finite $T$ since no partial spin
polarization is possible in 2DEG at $T=0$ \cite{att,dha1}.
However, in reference \cite{dha1} the authors find partial spin
polarization for $T/T_{F}$ between 0.3 and 1.6, i.e. well above
$T/T_{F}=0.02$ reported in reference \cite{ghosh}. On the other
hand, $r_{s}=7.6$ is considerably lower than the lowest value for
spin polarization in 2DEG \cite{ceper}.
\\

\section{\label{sec:conc}Summary}

 In summary, we have calculated the ferromagnetic critical density
at $T=0$ of the quasi-two-dimensional electron gas confined in
semiconductor GaAs-based symmetrically-doped quantum wells. We use
the screened Hartree-Fock approximation and prove rigorously that
the pronounced decrease of the transition density with the well
width is governed by the interplay between the in-plane and the
out-of-plane components of the exchange interaction. The
combination of these exchange terms with the Thomas-Fermi
screening produces a universal upper value for the well width
beyond which the polarized state cannot exist. We add different
effective masses for both spin polarizations, which are introduced
in order to take into account Coulomb correlations beyond
screening. Once the value of the effective mass for the polarized
phase is adjusted so as to reproduce the transition density for
the pure 2D case calculated with the VMC method, our theory gives
ground-state energy shifts that agree with those calculated within
this method. On the other hand, a density-dependent ratio between
both effective masses is required to fit the ground-state energy
shifts calculated with the FN-DMC method. Based on our theory and
the existing MC calculations for the 2DEG, we predict that narrow
quantum wells (with well widths roughly in the range 30\AA $\leq
d_{W}\leq$ 50\AA) should exhibit a ferromagnetic transition at a
density range between $N_{sc}=3.2\times10^{7}$cm$^{-2}$
($r_{s}\approx 100$) and $N_{sc}=6.1\times10^{8}$cm$^{-2}$
($r_{s}\approx 23$). This range, which looks far from the
densities achievable nowadays in GaAs quasi-2DEG, is already
within reach in GaAs quasi-2DHG systems.

\begin{acknowledgments}
The authors acknowledge partial support from Proyectos UBACyT
2001-2003 and 2004-2007, ANPCyT project PICT 19983, and
Fundaci\'{o}n Antorchas. P.I.T.\ is a researcher of CONICET.
\end{acknowledgments}


\begin{thebibliography}{99}
\bibitem{com1} We observe a missing factor $4\pi$ in equations (12), (13) and (14)
 of reference \cite{pk} and an incorrect $k$ factor in equation (14).
\end{thebibliography}
\end {document}